\documentclass[12pt]{article}
\usepackage[dvipdfm]{color}
\usepackage{graphics}
\usepackage{graphicx}
\usepackage{amssymb}
\usepackage{cite}
\newcommand{\Eqn}[1]{&\hspace{-0.5em}#1\hspace{-0.5em}&}

\def\ep{\epsilon}

\def\pint#1 {- \!\!\!\!\!\!\!\! \,\int_{#1}}

\def\ni       {\noindent}

\def\lbb     {\left[ }
\def\rbb      {\right] }
\def\comma      { \, , }
\def\period     { \, . }
\def\semiket#1  { \, #1 \, \rangle \, }
\def\del        {  \partial  }

\def\half       {  {1\over 2}  }
\def\abs#1      {  \, \vert #1 \vert \,   }
\def\Im#1    { \, {\rm Im } \, #1  }
\def\Re#1    { \, {\rm Re}  \, #1  }

\def\binom#1#2 { \vecii{ {}_{#1} }{\raisebox{.5ex}{$ {}^{#2} $}} }
\def\sqbinom#1#2 { \Bigl(\begin{array}{c} {}_{#1}
                       \\ \raisebox{.5ex}{${}^{#2}$} \end{array}\Bigr)^2  }

\def\siml    { \ \raisebox{-.6ex}{$ \stackrel{\!<}{\sim} $} \ }

\def\sql    {\sqrt{\lambda}}
\def\r12    {\frac{r_1}{r_2}}

\def\calO  {{\cal O}}
\def\calN  {{\cal N}}

\def\tilb   {\tilde{b}}

\def\tily   {\tilde{y}}
\def\tilx   {\tilde{x}}
\def\till   {\tilde{l}}
\def\tilf   {\tilde{f}}
\def\tildel   {\tilde{\delta}}

\def\vecii#1#2      {  { #1 \choose #2 }  }
\def\veciii#1#2#3   {  \left(\begin{array}{c}#1\\#2\\#3\end{array}\right)  }
\def\matrixii#1#2#3#4            {  \Bigl( \begin{array}{cc}#1&#2\\#3&#4
                                       \end{array} \Bigr) }
\def\matrixiii#1#2#3#4#5#6#7#8#9 {  \left(\begin{array}{ccc}#1&#2&#3\\
                                     #4&#5&#6\\#7&#8&#9\end{array}\right)  }
\def\eqb         {  \begin{eqnarray}  }
\def\eqe           {  \end{eqnarray}  }
\def\nn               {  \nonumber  }

\def\msection#1{ \addtocounter{section}{1} \setcounter{subsection}{0}
   \par \bigskip
      \par \bigskip \noindent
   {\bf \arabic{section} \quad  #1 }
    \par \bigskip}
\def\appsection#1{\addtocounter{section}{1} \setcounter{subsection}{0}
      \par \bigskip \par \bigskip \noindent
   {\bf \Alph{section} \quad  #1 }
    \par \bigskip}

\def\csectionast#1    { \begin{center}
    {\large\bf #1  }   \end{center} \par \bigskip}
\def\titleandfile#1#2   {  \begin{center}{\large\bf #1}\end{center}
                            \par\begin{flushright} #2 \end{flushright} 
                            \par \begin{flushright} \today \end{flushright}}

\renewcommand{\thefootnote}{\fnsymbol{footnote}}

\begin{document}
%
\def\papertitlepage{\baselineskip 3.5ex \thispagestyle{empty}}
\def\preprinumber#1#2#3{\hfill \begin{minipage}{3cm}  #1
              \par\noindent #2
              \par\noindent #3
             \end{minipage}}
\renewcommand{\thefootnote}{\fnsymbol{footnote}}
%
%
\papertitlepage
\setcounter{page}{0}
\preprinumber{}{UTHEP-522}{hep-th/0607190}
\baselineskip 0.8cm
\vspace*{2.0cm}
\begin{center}
{\large\bf\mathversion{bold}
A large spin limit of strings on $AdS_5 \times S^5$ \\
in a non-compact sector}
\end{center}
\vskip 4ex
\baselineskip 1.0cm
\begin{center}
        { Kazuhiro ~Sakai\footnote[2]{\tt sakai@phys-h.keio.ac.jp}, } \\
 \vskip -1ex
    {\it Department of Physics, Keio University}
 \vskip -2ex   
    {\it Hiyoshi, Yokohama 223-8521, Japan} \\
 
 \vskip 2ex
     { Yuji  ~Satoh\footnote[3]{\tt ysatoh@het.ph.tsukuba.ac.jp}}  \\
 \vskip -1ex
    {\it Institute of Physics, University of Tsukuba} \\
 \vskip -2ex
   {\it Tsukuba, Ibaraki 305-8571, Japan}

\end{center}
\vskip 13ex
%
\baselineskip=3.5ex
\begin{center} {\large\bf Abstract} \end{center}
We study the scaling law of the energy spectrum
of classical strings on $AdS_5 \times S^5$, in particular,
in the $SL(2)$ sector for large $S$ ($AdS$ spin) 
and fixed $J$ ($S^1 \subset S^5$ spin).
For any finite gap solution, we identify the limit in which the energy
exhibits the logarithmic scaling in $S$, characteristic to
the anomalous dimension of low-twist gauge theory operators.
Our result therefore shows that the $\log S$ scaling,
first observed by Gubser, Klebanov and Polyakov
for the folded string, is universal also on the string side,
suggesting another interesting window to explore 
the AdS/CFT correspondence as in the BMN/Frolov-Tseytlin limit.

\vskip 2ex
%
%
%
%
%
\vspace*{\fill}
\ni
July 2006
\newpage
\renewcommand{\thefootnote}{\arabic{footnote}}
\setcounter{footnote}{0}
\setcounter{section}{0}
\baselineskip = 3.3ex
\pagestyle{plain}
%
%
%
\msection{Introduction}

The integrable structures underlying in the $\calN = 4$ super Yang-Mills (SYM) 
theory \cite{MZ,BS1}  and the string theory on $AdS_5 \times S^5$ \cite{BPR}
have enabled us to probe the gauge/gravity correspondence quantitatively 
beyond the supersymmetric sectors. At one-loop in the planar
limit on the SYM side, the scaling dimension of the long operators
is found to precisely match the energy
of the rotating strings with large spins
\cite{FT1,MBSZ,BFST,Kruczenski1,KMMZ,Tseytlin,Beisert1}. 
In particular, in that (Frolov-Tseytlin) limit \cite{FT1,Tseytlin}, 
the scaling dimension/energy universally exhibits the BMN scaling \cite{BMN}. 
A systematic way to show the matching between the two sides is 
to compare the algebraic curves and the differentials 
thereon which are associated with the integrability 
\cite{KMMZ,Beisert1,KZ,BKS,Nameki,BKSZ1,BKSZ2,GKSV}.

In the non-compact sectors, there is another interesting large charge limit:
from the point of view on the string side,
it is  $S$ ($AdS_{5}$ spin) $ \gg 1$ 
with $J$ ($S^{5}$ spin) fixed, whereas, on the gauge theory side, 
it corresponds to considering low-twist ($J$) operators with large
Lorentz spin ($S$) \cite{GKP,FT2}. In this case, 
the anomalous dimension on the gauge theory side universally behaves 
as $\Delta -S \sim c\lambda  \log S$ at one loop, 
where $\lambda$ is the 't Hooft coupling 
and $c$ is a constant  \cite{CG,BGK1,Korchemsky,BGK2}. 
On the string theory side, some classical solutions with   
this $\log S$ scaling are found \cite{GKP,FT2,Ryang,Kruczenski2}
(see also \cite{HN}),
but with different dependence on $\lambda$.

The non-compact sectors are of interest, 
since any gauge theories, including QCD, possess such sectors and 
integrable structures emerge rather ubiquitously (see, e.g.,\cite{BGK1,Lipatov,FK}).
The anomalous dimension of low-twist operators, accounting for
the violation of the Bjorken scaling, is also related at large $S$ to the cusp
anomalous dimension and, hence, to certain physical processes.
Moreover, the analysis in the large $S$ limit with fixed $J$ may provide us
important data for the higher-loop Bethe ans\"atze
\cite{BDS,AFS,Staudacher1,BS2,Beisert2,Janik,Beisert3},
which have been studied mainly in the large $J$ limit.

In this note, we consider the finite gap solution on the string side
in the $SL(2)$ sector for large $S$ and fixed $J$. Related 
discussions on both string and gauge theory sides are found 
in \cite{BGK1,Korchemsky,BGK2,Lipatov,FK,ES}.
In particular, detailed analyses of the spectrum of the gauge theory operators 
and of the higher-loop Bethe ans\"atze are given in \cite{BGK2} and \cite{ES}, 
respectively. 
In terms of the algebraic curve, we identify the limit in which the energy of 
any finite gap solution of the string sigma model takes the form 
$\Delta-S \sim c \sqrt{\lambda} \log S$, where $\sql$ is the effective 
string tension and $c$ is a constant.
Therefore, the $\log S$ scaling, which is characteristic to the anomalous 
dimension of the low-twist gauge theory operators,  is universal also on the
string side, as the BMN scaling in the BMN/Frolov-Tseytlin limit.
This suggests another interesting window to explore the AdS/CFT 
correspondence.

In the next section, following \cite{KZ}, 
we summarize the finite gap solution in the $SL(2)$ sector.
In section 3, we analyze the general two-cut solution and 
identify the conditions under which the $\log S$ scaling emerges.
In the course, we explicitly write down the period conditions for the
differential. We briefly summarize the results of the corresponding 
integrals in Appendix. In section 4, based on the results in section 3,
we discuss the general finite gap solution. We show that, 
in a certain limit with $S/\sql \gg 1$ and  fixed $J/\sql$, 
the energy of any finite gap solution exhibits the $\log S$ scaling,
but with different dependence on $\lambda$ from the
perturbative gauge theory case.

\msection{Classical strings in the $SL(2)$ sector}

Let us consider a classical string in the $SL(2)$ sector,
namely, a string moving in $AdS_3 \times S^1$
in $AdS_5 \times S^5$ \cite{KZ}.
One can choose a gauge in which the coordinate 
field of $S^1$ takes the form
\eqb
   \phi = l \tau + m \sigma \comma
\eqe
where $\tau, \sigma$ are the  world-sheet coordinates.
Here both $\phi$ and $\sigma$ are periodically identified
$\phi\cong\phi+2\pi, \ \sigma\cong\sigma+2\pi$,
so that $m $ is an integer.
In terms of the angular momentum ($J$) associated with $S^1$ and
the effective string tension ($\sql$), $l$ is given by $l = J/\sql$. In this gauge, 
only the $AdS_3$ part is thus non-trivial. For a class of solutions called 
the finite gap solution, the motion of the string 
is specified by a differential 
\eqb
   dp = -\pi \frac{dx}{y} \Bigl[ l_+ f_+(x) + l_- f_-(x) 
  + \sum_{k=1}^{K-1} b_k x^{k-1} \Bigr]
\eqe
on a hyperelliptic curve given by
\eqb
   y^2 = \prod_{a=1}^{2K} (x-x_a) \period
\eqe   
Here, $x_a \in \bf R$,  $b_k$ are constant, $l_\pm = l \pm m$, and 
\eqb
   f_\pm (x) =  \frac{y_\pm}{(x \mp 1)^2} + \frac{y'_\pm}{x \mp1} 
\eqe
with $y_\pm = y \vert_{x = \pm 1}$ and 
$y'_\pm = \del_x y \vert_{x = \pm 1}$. The differential has to satisfy 
the following period conditions:
\eqb
    \oint_{A_a} dp =0 \comma \qquad \int_{B_a} dp = 2\pi n_a \period 
    \label{periodcond}
\eqe
The contours $A_a$ surround the $K-1$ (among $K$) cuts, whereas $B_a$
traverse each cut and terminate at the infinity
on each of the two sheets (see Fig.1).
The full $AdS_3$ sigma model actually possesses two kinds of $SL(2)$
excitation modes associated with the left and right multiplication
of the $SL(2)$ group coordinates.
For the sake of comparison to the minimal non-compact sector
on the gauge side, it is enough to take $ |x_a| > 1$
so that excitations in a single $SL(2)$ are present \cite{BKSZ1}.
%
\begin{figure}[ht]
   \hspace*{9ex}
    \includegraphics*[width=12cm]{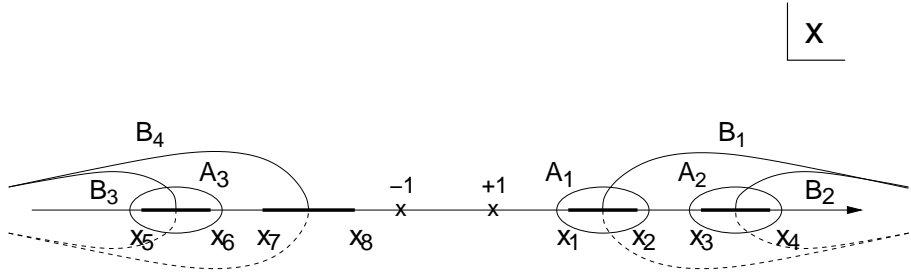}
  \caption{Contours $A_a$ and $B_a$ for $K=4$. The dashed lines represent
 the part of the contours on the other Riemann sheet.}
  \vspace*{1ex}
\end{figure}
The asymptotic behaviors of $dp$ for $x \to \infty$ and $x \to 0$ 
give relations to the energy $\Delta$ and the $AdS$ spin $S$:
\eqb
   \frac{2}{\sql}(\Delta + S) \Eqn{=}  b_{K-1} \comma \nn \\
  \frac{2}{\sql}(\Delta - S) \Eqn{=} \frac{1}{y(0)} \lbb l_+ (y_+ -y'_+) 
  +\ l_- (y_- +y'_-) + b_1 \rbb \period \label{asympt}
\eqe

\vspace*{-1ex}
\msection{General two-cut solutions}

As a preliminary to the discussion on the general finite gap 
solution, we first consider the general two-cut solution.
A discussion on the symmetric case is found in \cite{BGK2}. 

In this case, there are three independent period conditions.
It is possible to carry out their integrals explicitly. The results are
summarized in Appendix. One can then
combine those period conditions, to obtain a simpler set: 
\eqb
  2 b_1 \Eqn{=} - 2\Bigl(\frac{l_+}{y_+}-\frac{l_-}{y_-}\Bigr)x_0
                 + \Bigl(\frac{l_+}{y_+}+\frac{l_-}{y_-}\Bigr)
      \left[(x_1x_2+x_3x_4+2)-x_{13}x_{24} \frac{E(1-k)}{K(1-k)} \right] \comma \nn \\
n_1-n_2\Eqn{=}\frac{\pi}{2}\frac{\sqrt{x_{13}x_{24}}}{K(1-k)} 
  \Bigl(\frac{l_+}{y_+}+\frac{l_-}{y_-}\Bigr)  \comma \label{2cutperiod}\\
   n_1 \Eqn{=}  \Bigl(\frac{l_+}{y_+}+\frac{l_-}{y_-}\Bigr)
    \sqrt{x_{13}x_{24}}    
       G\Bigl( \sqrt{ \frac{ x_{13} }{ x_{23} }} \,, k \Bigr) 
       - \frac{l_+}{y_+} (x_{1} -1)  - \frac{l_-}{y_-} (x_{1} +1)  \comma \nn
\eqe
where $x_{ab} = x_a - x_b$, $ 2x_0 = \sum_{a=1}^4 x_a $, 
$k = x_{14}x_{23}/x_{13}x_{24}$, 
\eqb
    G(z,k) \equiv E(z,k) + F(z,k)\Bigl( \frac{E(1-k)}{K(1-k)} -1 \Bigr)  \comma
\eqe
and $E(k),K(k), E(z,k), F(z,k)$
denote standard elliptic integrals, our conventions of
which are found in Appendix.
$n_2$ is expressed similarly  to $n_1$, but that is not
independent of the above. In the case of the symmetric cuts, i.e., 
$x_1 = -x_4,\ x_2 = -x_3$, these conditions reduce to those in \cite{BGK2,SS0}.

Now, we would like to consider the case when 
\eqb
  \gamma \equiv \Delta - J - S
\eqe
 scales as $\log S$ for large $S/\sql$  and fixed $J/\sql, m$. 
 In the following, we focus on the case in which one cut is on the left side of $x=0$,
and the other is on the right side, i.e., $x_3 < x_4 < -1$, $1 < x_1 < x_2$.
The analysis of the case in which both cuts are on the same side of the origin 
is similar. For our purpose, we first note that
(\ref{asympt}) with $K=2$ gives an expression of $\gamma $ 
in terms of $x_a$:
\eqb
  && \frac{2}{\sql} (\gamma + J) \Bigl(1 - \frac{1}{\sqrt{\prod x_{a}} } \Bigl) 
   \ =  \,  \frac{4S/\sql}{\sqrt{\prod x_{a}} }
    \label{gammaxa} \\
  &&    \qquad + \,  l_+ \sqrt{\prod\frac{x_a -1}{x_{a}} } 
  \Bigl( 1+ \half \sum \frac{1}{x_{a} - 1} \Bigr)
       +  l_- \sqrt{\prod \frac{x_a+1}{x_{a}} } 
  \Bigl( 1- \half \sum \frac{1}{x_{a} + 1} \Bigr)
         \period \nn
\eqe
Second, from the first equation in (\ref{asympt}), it follows that 
\eqb
    \sql b_1 = \calO\Bigl( {\max}(\gamma, S) \Bigr) \period \label{b1}
\eqe    
Third, 
the first condition in (\ref{2cutperiod}) then implies that, for $S$ to be large,
at least one of the following conditions has to be satisfied:  
\eqb
   x_{2} \gg 1 \comma \quad |x_{3}| \gg 1 \comma \quad 
   x_{1}-1 \ll 1 \comma \quad |x_{4} +1| \ll 1 \period      
\eqe 

With these in mind, one can analyze the spectrum of $\gamma$ for large $S$.
We omit details
but, after some analysis, we find that $\gamma \sim c \sql \log (S/\sql) $
for large $S/\sql$ and fixed $ l, m$ with $c$ a constant independent of $J$,
when the following two conditions are satisfied:
\par\vspace*{1.8ex}\ni
\quad \ $1$.  \, 
    the inner branch points $x_1, x_4$ are of $\calO(1)$, and at least 
   one of them approaches the singularities of $dp$ at $x = \pm 1$, 
\par\medskip\ni
\quad \ $2$.  \, 
    the mode numbers $n_1, n_2$ are of $\calO(1)$. 
\par\vspace*{1.8ex}\ni
Throughout in this paper relation by tilde ($\sim$) signifies equality in
the leading order approximation.  
One can check that these conditions are equivalent to condition $1$ and 
\par\vspace*{1.8ex}\ni
\quad \ $2'$.  \,  $b_1, x_2, x_3  = \calO(S/\sql)$.
\par\vspace*{1.8ex}\ni
(See also the next section.)
These are consistent with the result for the symmetric two-cut case \cite{BGK2,SS0}. 
In addition, the independence from $J$ is in accord with
the fact on the gauge theory side that the lowest anomalous dimension, which is
described by a hyperelliptic curve with degenerate two cuts, is 
independent of the length of the operators \cite{BGK2}. 
  
We however remark that, 
in general, large $S$ does not necessarily mean the logarithmic scaling, but  
various asymptotics of $\gamma$ for large $S$ are allowed classically
by adjusting the parameters. To understand the difference from the gauge 
theory side, which always 
gives the $\log S$ scaling, it would be useful to note the allowed
values of the mode numbers $n_a$: on the string side they can be
large, whereas those on the gauge theory side are bound by $J$.\footnote{
This bound may emerge also on the string side if one
imposes some `quantization condition' respecting the integrality of $J$.}
Condition 2 eliminates such large mode numbers. In addition, our
analysis here is different from that for the symmetric two-cut case on
the string side in \cite{BGK2}, because the asymmetry of cuts, in addition to
large mode numbers, is allowed. In fact, one can check that 
$\gamma = \calO(S)$ with $n_1 = \calO\bigl( S/\sql \bigr), n_2 = \calO\bigl( (S/\sql)^{2} \bigr)$ 
when $ x_{1,2} = \calO(S/\sql)$ and $x_3 = -1 -\calO\bigl( (S/\sql)^{-1} \bigr), 
x_4 = -1 - \calO\bigl( (S/\sql)^{-3} \bigr)$.

 A concrete example to give the $\log S$ scaling
is the case in which $ |x_{2,3}| \gg 1$ and 
$ x_{1}= 1 +\delta_{1}^{2},  x_{4} = -1 - \delta_{4}^{2}$ with 
$\delta_{1}, \delta_{4} \ll 1$.
In this case, $k \sim 2 x_{32}/x_{2}x_{3} \ll 1$. Thus,  
using the asymptotic forms of the elliptic integrals in (\ref{smallk}), 
the period conditions are reduced to
\eqb
   \frac{\sqrt{2}}{\log(1/k)} \Bigl( \frac{l_{+}}{\delta_{1}} 
  + \frac{l_{-}}{\delta_{4}} \Bigr)
   \sim \frac{n_{1}}{ \arcsin \sqrt{\frac{ -x_{3}}{x_{23} } }} 
   \sim \frac{-n_{2}}{ \arcsin \sqrt{\frac{x_{2}}{x_{23} } }} 
   \sim \frac{2 b_{1}}{\sqrt{-x_{2}x_{3}}} \period \label{2cutperiods}
\eqe
Since $n_{1,2} = \calO(1)$, the above relation between
$b_{1}$ and $n_{1,2}$ implies that  $x_{2,3} = \calO(b_{1})$. 
It turns out  that $b_{1}, x_{2,3} = \calO(S/\sql)$ for (\ref{b1}) to be satisfied.
Together with (\ref{gammaxa}), the relation between $\delta_{1,4}$ and $n_{1}$
then gives 
 \eqb
    \gamma \sim \frac{ \sql n_{1}}{4 \arcsin \sqrt{\frac{ -x_{3}}{x_{23} } } } 
\log \frac{S}{\sql}
     \period
 \eqe
The relation between $n_{1}$ and $n_{2}$ constrains the coefficient
 of $\log S$. 
Taking into account this, one finds that, 
when $n_{1} = -n_{2} = 1$ and $ x_{2} = -x_{3} $,  $\gamma$ takes the minimum
$\gamma_{\rm min} = (\sql/\pi) \log (S/\sql)$, which agrees with the energy of 
the folded
strings corresponding to the symmetric two-cut solutions \cite{GKP,FT2}.

%
\msection{General finite gap solutions}

In this section, we generalize the analysis in the previous section to 
the case of the general finite gap solution. We show that
there exists a limit in which the energy of any finite gap solution
for large $S/\sql$ and fixed $J/\sql,m$
behaves as $\gamma \sim c \sql \log (S/\sql)$ with
$c$ a constant of order 1.

Here, we would like to emphasize that our point is to show a
sector-wise correspondence to the gauge theory for large $S$, as in
the Frolov-Tseytlin limit \cite{Kruczenski1,KMMZ,KZ,BKS,Nameki,BKSZ1,BKSZ2}: 
The $\log S$ scaling holds not only
for multi-soliton solutions, whose variation from the ground state is
relatively small, but also for general quasi-periodic solutions, which
could be far apart from the ground state. Such a sector-wise
correspondence would also be useful to study the correspondence of the
operators/solutions between the gauge/string sides, which is generally
quite non-trivial (e.g., \cite{HM}). Moreover, our results
may give useful insights into the asymptotic Bethe ans{\"a}tze \cite{BDS,AFS,Staudacher1,BS2,Beisert2,Janik,Beisert3},
which are expected to interpolate all the states between the
gauge/string sides.

For this purpose, it is useful to introduce rescaled variables
\eqb
    \tilx = \frac{x}{M} \comma \quad \tilx_a = \frac{x_a}{M} \comma \quad
    \till_\pm = \frac{l_\pm}{M} \comma \quad \tilb_k = \frac{b_k}{M^{K-k}} \period
\eqe
In terms of these, the differential $dp$ becomes
\eqb
     dp = -\pi \frac{d\tilx}{\tily} \Bigl[ \till_+ \tilf_+(\tilx) 
  + \till_- \tilf_-(\tilx) 
  + \sum_{k=1}^{K-1} \tilb_k \tilx^{k-1} \Bigr] \comma
\eqe
where $ \tily^2 = \prod_{a=1}^{2K} (\tilx-\tilx_a)$, 
\eqb
   \tilf_\pm (x) =  \frac{\tily_\pm}{(\tilx \mp \ep)^2} 
  + \frac{\tily'_\pm}{\tilx \mp \ep}  \comma 
\eqe
$ \tily_\pm = \tily \vert_{\tilx = \pm \ep} \comma 
      \tily'_\pm = \del_{\tilx} \tily \vert_{\tilx = \pm \ep} $, 
and $\ep = 1/M$. From (\ref{asympt}),  one finds that
$ \frac{2}{\sql}(\Delta + S) =  M \tilb_{K-1}$ and
\eqb
   \frac{2}{\sql}(\Delta - S) 
   = \frac{1}{\tily(0)} \lbb \till_+ \tily_+ 
   \Bigl( \frac{1}{\ep} + \half \sum \frac{1}{\tilx_a -\ep} \Bigr)
   +  \till_- \tily_- 
   \Bigl( \frac{1}{\ep} - \half \sum \frac{1}{\tilx_a +\ep} \Bigr)
  +  \ep \tilb_1 \rbb \period 
   \label{D-S}
\eqe 

When 
\eqb
    l = \frac{J}{\sql}, \, \frac{S}{\sql} = \calO\bigl(M\bigr)
     \gg 1 \comma \quad  
   m, \tilb_k, \tilx_a = \calO(1) \comma
    \label{FT}
\eqe      
the period conditions (\ref{periodcond}) give
relations among quantities of order 1; nothing is very large or small (generically).
$\tilb_k, \tilx_{a}$ of order 1 solve these conditions for the winding number $m$, 
mode numbers and fillings (i.e., the contributions to $S$ from each cut) of 
order 1, giving a classical string solution. 
The energy is then obtained by expanding (\ref{D-S}) up to and including terms 
of $\calO(\ep)$ and setting $\ep = 1/4\pi l$:
\eqb
  \gamma \sim  \frac{\lambda}{128\pi^{2} J} \Bigl(  \sum_{a} \frac{1}{\tilx_{a}^{2}} 
    - \sum_{a>b} \frac{2}{\tilx_{a}\tilx_{b}}  
    + \frac{16 \pi \tilb_{1}}{ \tily(0) }\Bigr) \period \label{BMN} 
\eqe
Since $ 2\pi \tilb_{1} \sim 1 + 2 S/J$, this $\gamma$ precisely 
agrees with 
the gauge theory result in the Frolov-Tseytlin limit (cf. Eq. (3.13) in \cite{KMMZ}), 
to confirm  the matching between the string and the one-loop gauge theory results
shown in \cite{KZ}. 

In the following, based on the result in the previous section, we consider the case
in which $K \geq 2$, 
\eqb
   M= \frac{S}{\sql} \gg 1 \comma \quad l, m, \tilb_{k} = \calO(1)
   \comma  \quad  \tilx_a = \calO(1) \ (a \neq 1,2K) \comma \label{cond1}
\eqe   
the innermost branch points $\tilx_1, \tilx_{2K} $ are of
$\calO(\ep)$ and at least one of them approaches $\pm \ep$. 
Thus, we denote the innermost branch points by 
\eqb
    \tilx_1 = \ep +\tildel_1^2 \comma \qquad \tilx_{2K} = -\ep -\tildel_{2K}^2 
     \comma \label{cond2}
\eqe
where $\tildel^{2}_{1,2K} \siml \ep$ and 
at least one of $\tildel^{2}_{1,2K} \ll \ep$. We also consider the generic case
so that $\tilx_a \ (a\neq 1, 2K)$ do not collide each other, or the algebraic curve 
does not degenerate there.

Let us first analyze the $A_{1}$-period condition for one of the innermost cuts. 
To evaluate the integral, we note that, for $ \tilx \in (\tilx_1, \tilx_2) $,
\eqb 
   \tily \sim c \tily^{(2)} 
    \comma \qquad \sum_{k=1} \tilb_{k} \tilx^{k-1} \sim c' \period
\eqe
Here and in the following, we denote by $c, c'$ numbers of order 1 and,
 by the superscript $(2)$, quantities for the two-cut case corresponding 
 to the two innermost cuts. For example, 
 $(\tily^{(2)})^{2} = (\tilx -\tilx_{1})(\tilx -\tilx_{2})(\tilx -\tilx_{2K-1})(\tilx -\tilx_{2K})$.
 In addition, $\tily_{\pm} \sim c \tily^{(2)}_{\pm}$, 
 $\tily'_{\pm}/\tily_{\pm}  \sim \tily^{(2)'}_{\pm}/\tily^{(2)}_{\pm} + c$ and, hence,
\eqb
   \tilf_{\pm}(\tilx) \sim  c \tilf_{\pm}^{(2)}(\tilx) 
  + \frac{c' \tily_{\pm}^{(2)}}{\tilx \mp \ep}
    \period
\eqe
 It turns out that the difference of $\tilf_{\pm}$ and $\tilf_{\pm}^{(2)}$ 
(the second terms 
 on the right-hand side) gives terms of order $\ep$ in the integral.
 The integral involving $\tilf_{\pm}^{(2)}$  can be read off 
from (\ref{Aperiod}).
 Retaining the terms relevant under the conditions (\ref{cond1}) and (\ref{cond2}), 
 the $A_{1}$-period condition becomes
\eqb
     0 \sim K(1-k) -  \Bigl( \frac{c\till_{+}}{\tily^{(2)}_{+}} 
     + \frac{c'\till_{-}}{\tily^{(2)}_{-}}\Bigr) \sqrt{\tilx_{13}\tilx_{24}} E(1-k) 
\period
\eqe
From the asymptotic behaviors of the elliptic integrals (\ref{smallk}), 
it then follows that
\eqb
    \frac{c\till_{+}}{\tily^{(2)}_{+}} 
     + \frac{c'\till_{-}}{\tily^{(2)}_{-}} \sim  \log \frac{S}{\sql} \comma
\eqe
and that min$(\tilde{\delta}^{2}_{1}, \tilde{\delta}^{2}_{2K}) \sim c\ep/\log^{2} \ep$.

It is straightforward to take into account other period conditions. 
We first note that $\till_\pm \tilf_\pm$ in $dp$ are neglected in the integrals,  
since they give contributions of order $ \ep^{\frac{3}{2} }/\tilde{\delta}_{1,2K} \ll 1$ 
and hence subleading to others.
Also, for these period conditions, $\tilx_{1,2K}$ can be set
to zero at the leading approximation, since $\tilx - \tilx_{1,2K} \sim \tilx$  
in evaluating the integrals. 
Thus, these conditions give relations among order 1 quantities, namely, 
$\tilb_{k}, \tilx_{a} (a \neq 1,2K) = \calO(1)$ and
\eqb
    n_a = \calO(1)\period
\eqe

We are now ready to estimate $\gamma$. 
In the present case, 
the dominant contribution to $\gamma$ comes from the terms
with $1/(\tilx_{1} -\ep)$ or $1/(\tilx_{2K} + \ep)$. Therefore,
\eqb
    \gamma \Eqn{\sim} \frac{\sql}{2\tily(0)} 
  \Bigl(  \frac{\till_{+} \tily_{+}}{\tilx_{1} -\ep}
    + \frac{\till_{-} \tily_{-}}{\tilx_{2K} +\ep}  \Bigr) \sim 
    c' \sql \Bigl( \frac{\till_{+}}{\tily^{(2)}_{+}} 
  + \frac{\till_{-}}{\tily^{(2)}_{-}}\Bigr) 
     \nn \\
    \Eqn{\sim} c \sql \log \frac{S}{\sql} \period 
\eqe

\par\bigskip

In sum, we have
shown that, in the limit given by (\ref{cond1}) and (\ref{cond2}),
the energy of the classical string scales as $ \gamma \sim c \sql \log S$ with $c $ 
a constant, generalizing the results in \cite{GKP,FT2,Ryang,Kruczenski2}.
Therefore, the $\log S$ scaling, which is characteristic to the anomalous 
dimension of the low-twist gauge theory operators, is universal also 
for the classical string in the $SL(2)$ sector, as the BMN scaling 
in the BMN/Frolov-Tseytlin limit (\ref{FT}).
This suggests another interesting window to explore the AdS/CFT 
correspondence. Our analysis may be extended to other
non-compact sectors along the line of \cite{BKS,Nameki,BKSZ1,BKSZ2,GKSV}.  

\vspace{5ex}
\ni {\bf Note added:} \  
After this work was completed, we were informed by Sergey Frolov that
he, with Matthias Staudacher, reached the same conclusion by a different method.
%
\vspace{4ex}
\begin{center}
  {\bf Acknowledgments}
\end{center}
We would like to thank N.~Gromov, V.~Kazakov, S.~Lee, T.~Mateos,
K.~Mohri, M.~Staudacher, B.~Stefanski, P.~Vieira, and K.~Yoshida
for useful discussions and conversations.
We would also like to thank G.~Korchemsky for a useful discussion 
and informing us of part of the results in \cite{BGK2} prior to
publication.  Research of K.S. is supported partly by
the Keio Gijuku Academic Development Funds.
Y.S. is very grateful to the theoretical physics group 
at Imperial College London, where part of this work was done, 
for its warm hospitality.
\vspace*{2ex}
\setcounter{section}{0}
\appsection{Period conditions for general two-cut solutions}

In this appendix, we summarize the period conditions for the general two-cut case.
For the $A$-period condition, we first observe an identity
\eqb
 - y \left(\frac{y}{x\mp 1}\right)'
= y_\pm f_\pm (x) +(x_0\mp 2)(x\mp 1)-(x\mp 1)^2 \comma \label{yiden}
\eqe
where $ 2 x_0 = \sum_{a=1}^4 x_a$. The A-period condition is then found to be
\eqb
  0 \Eqn{=} \frac{1}{2\pi i} \oint_A dp
= \frac{2}{2\pi i } \int_{x_1}^{x_2}\frac{dp}{dx}dx
    \nn \\
   \Eqn{=} \lbb \Bigl( \frac{l_+}{y_+} + \frac{l_-}{y_-} \Bigr) (x_1x_2+x_3x_4+2)
           -2\Bigl( \frac{l_+}{y_+} - \frac{l_-}{y_-} \Bigr)x_0 -2b_1 \rbb 
        \frac{1}{\sqrt{x_{13}x_{24}}} K(1-k) \nn \\
    && \hspace{2em}
   -  \Bigl( \frac{l_+}{y_+} + \frac{l_-}{y_-} \Bigr) \sqrt{x_{13}x_{24}} E(1-k) 
  \comma \label{Aperiod}
\eqe
where $x_{ab} = x_a - x_b$, and $k = x_{14}x_{23}/x_{13}x_{24}$. 
$K(k), E(k)$ are the elliptic integrals, our conventions of which
are
\eqb
F(z,k)=\int_0^z\frac{dx}{\sqrt{(1-x^2)(1-kx^2)}},\quad
E(z,k)=\int_0^z dx\sqrt{\frac{1-kx^2}{1-x^2}} \comma \label{ellint}
\eqe
and
$K(k)= F(1,k)$, $E(k) = E(1,k)$. 

To evaluate the B-period conditions, we first make a change of 
variables $ u = 1/x$, and consider $\Bigl( \frac{\sqrt{Q(u)}}{u\mp1}\Bigr)'$
where $Q(u) = u^{4} y^{2} = \prod_{a=1}^{4}(1-x_{a}u)$.
Using identities similar to (\ref{yiden}),  we obtain
\eqb
      n_1 \Eqn{=} \frac{1}{2\pi}\int_{B_1} dp
 = \frac{2}{2\pi}\int_{x_2}^{\infty}\frac{dp}{dx}dx  \nn \\
   \Eqn{=} 
   \lbb \Bigl( \frac{l_+}{y_+} + \frac{l_-}{y_-} \Bigr) (x_1x_4+x_2x_3+2)
          - 2\Bigl( \frac{l_+}{y_+} - \frac{l_-}{y_-} \Bigr)x_0 -2b_1 \rbb 
        \frac{1}{\sqrt{x_{13}x_{24}}} F(\sqrt{ \frac{x_{13}}{x_{23} }}, k) \nn \\
    && \hspace{2em}
      + \Bigl( \frac{l_+}{y_+} + \frac{l_-}{y_-} \Bigr) 
  \sqrt{x_{13}x_{24}} E(\sqrt{\frac{ x_{13}}{x_{23} }}, k)
   - \Bigl( \frac{l_+}{y_+} + \frac{l_-}{y_-} \Bigr) x_1 
  + \Bigl( \frac{l_+}{y_+} - \frac{l_-}{y_-} \Bigr) \comma \nn \\
    - n_2 \Eqn{=}- \frac{1}{2\pi}\int_{B_2} dp
 =-\frac{2}{2\pi}\int_{x_1}^{-\infty}\frac{dp}{dx}dx  \\
   \Eqn{=} 
    \lbb \Bigl( \frac{l_+}{y_+} + \frac{l_-}{y_-} \Bigr) (x_1x_4+x_2x_3+2)
           -2\Bigl( \frac{l_+}{y_+} - \frac{l_-}{y_-} \Bigr)x_0 -2b_1 \rbb 
        \frac{1}{\sqrt{x_{13}x_{24}}} F(\sqrt{\frac{x_{24}}{x_{23}} }, k) \nn \\
    && \hspace{2em}
      + \Bigl( \frac{l_+}{y_+} + \frac{l_-}{y_-} \Bigr) 
  \sqrt{x_{13}x_{24}} E(\sqrt{\frac{x_{24}}{x_{23} }}, k)
   + \Bigl( \frac{l_+}{y_+} + \frac{l_-}{y_-} \Bigr) x_4 
  - \Bigl( \frac{l_+}{y_+} - \frac{l_-}{y_-} \Bigr) \period \nn 
\eqe
These are related by $(x,p,m,x_a) \leftrightarrow (-x,-p,-m, -x_{4-a+1})$. 

In the main text, we use the asymptotic behaviors for small $k$ such as
\eqb
   && E(1-k) \sim 1 \comma  \quad K(1-k) \sim \half \log(1/k) \comma \nn \\
   && E(z,k) \sim F(z,k) \sim \arcsin z \period \label{smallk} 
\eqe 
  
%
%
\def\thebibliography#1{\list
 {[\arabic{enumi}]}{\settowidth\labelwidth{[#1]}\leftmargin\labelwidth
  \advance\leftmargin\labelsep
  \usecounter{enumi}}
  \def\newblock{\hskip .11em plus .33em minus .07em}
  \sloppy\clubpenalty4000\widowpenalty4000
  \sfcode`\.=1000\relax}
 \let\endthebibliography=\endlist
%
%
\newpage
\begin{center}
 {\bf References}
\end{center}
\par \vspace*{-2ex}

%

%


\begin{thebibliography}{999}
\parskip=-2.5pt
%
%
\bibitem{MZ}    J.~A.~Minahan and K.~Zarembo,
        JHEP {\bf 0303} (2003) 013
         [hep-th/0212208].
%
\bibitem{BS1} N.~Beisert and M.~Staudacher,
  %
  Nucl.\ Phys.\ B {\bf 670} (2003) 439
  [hep-th/0307042].
%
\bibitem{BPR} I.~Bena, J.~Polchinski and R.~Roiban,
  %
  Phys.\ Rev.\ D {\bf 69} (2004) 046002
  [hep-th/0305116].
%
\bibitem{FT1} S.~Frolov and A.~A.~Tseytlin,
  %
  Nucl.\ Phys.\ B {\bf 668} (2003) 77
  [hep-th/0304255].
%
%
\bibitem{MBSZ} N.~Beisert, J.~A.~Minahan, M.~Staudacher and K.~Zarembo,
        JHEP {\bf 0309} (2003) 010
        [hep-th/0306139].
%
\bibitem{BFST} N.~Beisert, S.~Frolov, M.~Staudacher and A.~A.~Tseytlin,
         JHEP {\bf 0310} (2003) 037
         [hep-th/0308117].
%
\bibitem{Kruczenski1} M.~Kruczenski,
        Phys.\ Rev.\ Lett.\  {\bf 93} (2004) 161602
       [hep-th/0311203].
%
\bibitem{KMMZ} V.~A.~Kazakov, A.~Marshakov, J.~A.~Minahan and K.~Zarembo,
         JHEP {\bf 0405}, 024 (2004)
         [hep-th/0402207].
%
%
\bibitem{Tseytlin} A.~A.~Tseytlin,
      hep-th/0311139.
%
\bibitem{Beisert1}  N.~Beisert,
        Phys.\ Rept.\  {\bf 405} (2005) 1
       [hep-th/0407277].
%
%
\bibitem{BMN} 
        D.~Berenstein, J.~M.~Maldacena and H.~Nastase,
        JHEP {\bf 0204} (2002) 013
        [hep-th/0202021].
%
%
\bibitem{KZ} V.~A.~Kazakov and K.~Zarembo,
        JHEP {\bf 0410}, 060 (2004)
        [hep-th/0410105].
%
 \bibitem{BKS} N.~Beisert, V.~A.~Kazakov and K.~Sakai,
          Commun.\ Math.\ Phys.\  {\bf 263} (2006) 611
         [hep-th/0410253].
 %
 \bibitem{Nameki} S.~Sch\"afer-Nameki,
  Nucl.\ Phys.\ B {\bf 714} (2005) 3
  [hep-th/0412254].
 %
 \bibitem{BKSZ1} N.~Beisert, V.~A.~Kazakov, K.~Sakai and K.~Zarembo,
          Commun.\ Math.\ Phys.\  {\bf 263} (2006) 659
          [hep-th/0502226].
 %
 \bibitem{BKSZ2} N.~Beisert, V.~A.~Kazakov, K.~Sakai and K.~Zarembo,
           JHEP {\bf 0507} (2005) 030
           [hep-th/0503200].
 %
 \bibitem{GKSV} N.~Gromov, V.~Kazakov, K.~Sakai and P.~Vieira,
         hep-th/0603043.
%
\bibitem{GKP} S.~S.~Gubser, I.~R.~Klebanov and A.~M.~Polyakov,
        Nucl.\ Phys.\ B {\bf 636} (2002) 99 [hep-th/0204051].
%
\bibitem{FT2} S.~Frolov and A.~A.~Tseytlin,
         JHEP {\bf 0206} (2002) 007
        [hep-th/0204226].
%
%
\bibitem{CG}  C.~G.~Callan and D.~J.~Gross,
  %
  Phys.\ Rev.\ D {\bf 8} (1973) 4383.
%
\bibitem{BGK1} A.~V.~Belitsky, A.~S.~Gorsky and G.~P.~Korchemsky,
  %
  Nucl.\ Phys.\ B {\bf 667} (2003) 3
  [hep-th/0304028].
 %
 \bibitem{Korchemsky} G.~P.~Korchemsky,
  Nucl.\ Phys.\ B {\bf 462} (1996) 333
  [hep-th/9508025].
 %
 \bibitem{BGK2} A.~V.~Belitsky, A.~S.~Gorsky and G.~P.~Korchemsky,
         Nucl.\ Phys.\ B {\bf 748} (2006) 24
        [hep-th/0601112].
%
%
%
\bibitem{Ryang} S.~Ryang,
  JHEP {\bf 0404} (2004) 053
  [hep-th/0403180].
%
\bibitem{Kruczenski2}      M.~Kruczenski,
           JHEP {\bf 0508} (2005) 014
           [hep-th/0410226].
%
\bibitem{HN} S.~A.~Hartnoll and C.~Nunez,
  JHEP {\bf 0302} (2003) 049
  [hep-th/0210218].
%
 %
 \bibitem{Lipatov} L.~N.~Lipatov,
   JETP Lett. {\bf 59} (1994) 596
  [hep-th/9311037].
%
\bibitem{FK} L.~D.~Faddeev and G.~P.~Korchemsky,
  Phys.\ Lett.\ B {\bf 342} (1995) 311
  [hep-th/9404173].
%
%
%
%
\bibitem{BDS}   N.~Beisert, V.~Dippel and M.~Staudacher,
  %
  JHEP {\bf 0407} (2004) 075
  [hep-th/0405001].
%
\bibitem{AFS} G.~Arutyunov, S.~Frolov and M.~Staudacher,
  JHEP {\bf 0410} (2004) 016
  [hep-th/0406256].
%
\bibitem{Staudacher1} M.~Staudacher,
  JHEP {\bf 0505} (2005) 054
  [hep-th/0412188].
%
\bibitem{BS2} N.~Beisert and M.~Staudacher,
  Nucl.\ Phys.\ B {\bf 727} (2005) 1
  [hep-th/0504190].
%
\bibitem{Beisert2} N.~Beisert,
  hep-th/0511082.
%
\bibitem{Janik} R.~A.~Janik,
  Phys.\ Rev.\ D {\bf 73} (2006) 086006
  [hep-th/0603038].
%
\bibitem{Beisert3} N.~Beisert,
  %
  hep-th/0606214.

%
\bibitem{ES}  B.~Eden and M.~Staudacher,
   J.\ Stat.\ Mech.\  {\bf 0611} (2006) P014
  [hep-th/0603157].
%
%
 \bibitem{SS0} K. Sakai and Y. Satoh, unpublished.
%
\bibitem{HM}  D.~M.~Hofman and J.~M.~Maldacena,
  J.\ Phys.\ A {\bf 39} (2006) 13095
  [hep-th/0604135].
%
\end{thebibliography}
\end{document}